\documentclass[prb,showpacs]{revtex4}
\usepackage{graphicx}
\begin{document}
\title{Dissipative electron-phonon system photoexcited far from equilibrium} 
\author{Navinder Singh and N. Kumar}
\email{nkumar@rri.res.in}
\affiliation{Raman Research Institute, Bangalore 560080, India}
\begin{abstract}
We derive the steady-state electron distribution function for a semiconductor driven far 
from equilibrium by the inter-band photoexcitation assumed homogeneous over the nanoscale sample. Our analytical treatment is based on the generalization of a stochastic model known for a driven dissipative granular gas. The generalization is physically realizable in a semiconducting sample where electrons are injected into the conduction band by photoexcitation, and removed through the electron-hole recombination process at the bottom of the conduction band. Here the kinetics of the electron-electron and the electron-phonon (bath) scattering processes, as also the partitioning of the total energy in the inelastic collisions, are duly parametrized by certain rate constants. Our analytical results give the steady-state-energy distribution of the classical (non-degenerate) electron gas as function of the phonon (bath) temperature and the rates of injection (cw pump) and depletion (recombination). Interestingly, we obtain an accumulation of the electrons at the bottom of the conduction band in the form of a delta-function peak $-$ a non-equilibrium classical analogue of condensation. Our model is specially appropriate to a disordered, indirect band-gap, polar semiconducting sample where energy is the only state label, and the electron-phonon coupling is strong while the recombination rate is slow. A possible mechanism for the dissipative inelastic collisions between the electrons is also suggested.
\end{abstract}
\maketitle
\noindent
{\bf INTRODUCTION}\\
The kinetics of evolution of the electron distribution function for an electron-phonon system driven far from equilibrium by photo-excitation is of considerable current interest, experimentally [1] (the pump-probe experiments) and theoretically [2,3,4] (e.g., the two-temperature model.  In the following, we develop a general stochastic approach to this kinetic problem as an alternative to the commonly used two-temperature model. Most such studies have been for metallic nanometric particles. More specifically, we derive the steady-state electron distribution function for the  semiconducting sample driven far 
from equilibrium by the inter-band photoexcitation assumed homogeneous over the sample. Our analytical treatment is based on the generalization of a stochastic model known for a driven dissipative granular gas [5]. The generalization is physically realizable in a semiconducting sample where electrons are injected into the conduction band by photoexcitation and removed at the bottom of the conduction band through the electron-hole recombination process. Here the kinetics of the electron-electron and the electron-phonon (bath) scattering processes, as also the partitioning of the total energy in the inelastic collisions, are duly parametrized by certain rate constants. Our analytical results give the steady-state electron distribution function, and the mean energy of the classical non-equilibrium electron gas as function of the phonon (bath) temperature and the rates of injection (cw pump) and depletion (recombination). Interestingly, we obtain accumulation of the electrons at the bottom of the conduction band in the form of a delta-function peak $-$ a non-equilibrium classical analogue of condensation! Our model is specially appropriate to a disordered, indirect band-gap, polar semiconducting sample where energy is the only state label, and the electron-phonon coupling is strong while the recombination rate is slow. A possible mechanism for the dissipative inelastic binary collisions between electrons is also suggested.
\vspace{0.40cm}

\noindent
{\bf THE MODEL}\\
Let $n_e(E)dE$ be the number of electrons lying in the energy range $\pm dE/2$ centred about $E$ in the conduction band of a semiconducting sample of volume $\Omega$. The electron-electron collisions, assumed inelastic in general, are described by the process; $ E_i + E_i^\prime \longrightarrow E_f + E_f^\prime = \alpha (E_i + E_i^\prime)$ with $\alpha \leq 1$, in which the {\em tagged} electron of energy $E_i$ collides with another electron of energy $E_i^\prime$ lying in the energy shell $E_i^\prime\pm\frac{1}{2}\Delta E_i^\prime$, and is scattered to the final state $E_f$. The scattering rate for this inelastic process is taken to be $(1-f)\Gamma n(E_i^\prime) dE_i^\prime$. Similarly, the electron-phonon scattering rate is given by $f\Gamma n_{ph}(E_i^\prime) dE_i^\prime$, with $n_{ph}(E_i^\prime) dE_i^\prime$ as the number of thermal  phonons in the phonon-energy shell $E_i^\prime\pm\frac{1}{2}\Delta E_i^\prime$. Here, the fraction $0\leq f\leq 1$ determines the relative strengths of the binary electron-electron and the electron-phonon collisions. Also, let the electrons be injected through photo-excitation into the conduction band at energy $E_{ex}$ at the rate $g_{ex}\delta(E-E_{ex})$, and then be removed(depleted) from the conduction band at energy $E_d < E_{ex}$ through recombination. Here the phonons are assumed to remain in thermal equilibrium at temperatures $T$. We expect this depletion-by-recombination to occur mostly at the bottom of the conduction band for which $E_d \simeq 0$. This depletion rate can be modelled by a term $-g_d\delta(E-E_d)n_e(E_d)$. In our model sample we assume a uniform density of states for the electrons and the energy to be the only label for the single particle states. The photo-excitation is taken to be homogenous over the sample, which is reasonable for a nanoscale disordered semiconducting sample. For the above dissipative model driven far from equilibrium, the kinetics for the non-equilibrium electron number density $n_e(E)$ is given by the rate equation
\begin{eqnarray}
&&\frac{\partial n_e(E)}{\partial t}= -n_e(E)\int dE^\prime [n_e(E^\prime)(1-f)
+n_{ph}(E^\prime)f]\Gamma  \nonumber \\
&& +\int_0^1 dz p(z) \int dE^\prime \int dE^{\prime\prime} \delta(E-z\alpha (E^{\prime}+
E^{\prime\prime}))n_e(E^\prime)n_e(E^{\prime\prime})(1-f)\Gamma \nonumber \\
&&+\int_0^1 dz p(z) \int dE^\prime\int dE^{\prime\prime} \delta(E-z (E^{\prime}+
E^{\prime\prime}))n_e(E^{\prime})n_{ph}(E^{\prime\prime})f\Gamma \nonumber \\
&&+g_{ex}(t)\delta(E-E_{ex})-g_d\delta(E-E_d)n_e(E_d).
\end{eqnarray}

In the above, we have assumed the total energy $(E^{\prime}+E^{\prime\prime})$ for a binary 
collision to be partitioned such that a fraction $z$, with probability density $p(z)$, goes to the {\em tagged} electron of initial energy $E^{\prime}$, and $1-z$ to the colliding particle (electron or phonon of initial energy $E^{\prime\prime}$). The inclusion of $\alpha$ in the electron-electron collision takes care of the possibility of inelastic electron-electron collisions. Note that we have suppressed the time argument ($t$) in the non-equilibrium electron-number density $n_e(E)$. Taking the energy Laplace transform
\begin{equation}
\tilde{n}_e(s)=\int_0^\infty e^{-sE}n_e(E)dE,
\end{equation}
of Eq. (1), we obtain,
\begin{eqnarray}
\frac{\partial}{\partial t} {\tilde{n}}(s) &=& - \Gamma \tilde{n}_{e}(s)[(1-f)N_e+fN_{ph}]+(1-f)\Gamma \int_0^1 p(z) dz\,\,
\tilde{n}_{e}^2(\alpha z s)\nonumber \\
&&+f\Gamma \int_0^1 dz \,\,p(z)\,\,\tilde{n}_{e}(z s)\tilde{n}_{ph}(z s)\nonumber\\ && +g_{ex} (t) e^{-sE_{ex}}-g_d e^{-sE_d}n_{e}(E_d).
\end{eqnarray} 

In the following, we will consider for simplicity the steady-state condition under constant (cw) photoexcitation, $g_{ex}(t)=g_{ex}$. A pulsed excitation can, of course, be considered in general. Accordingly, we set $\frac{\partial}{\partial t} {\tilde {n}}_e (s) = 0$ above, and all quantities  on the R.H.S. of Eq.(3) are then independent of time. 

In order to calculate the  steady-state electron distribution for the system in terms of the bath (phonon) temperature and other rate parameters, we expand $\tilde{n}_e(s)$ in powers of the Laplace-transform parameter $s$ as
\begin{equation}
\tilde{n}_e(s)=N_e - s\langle E_e \rangle + s^2 \langle E_e^2 \rangle . . . ,
\end{equation}
and equate the co-efficients of like powers of $s$. Thus, from the  zeroth power of s, we obtain at once
\begin{equation}
n_e(E_d) = (g_{ex}/g_d).
\end{equation}

\noindent
Similarly, from the first power of s, we get,

\begin{equation}
\langle e_e \rangle=\frac{(f/2)\langle e_{ph}\rangle}{\rho_{e-ph}(1-\alpha)(1-f)+f/2}
+\frac{g_{ex}(E_{ex}-E_d)/\Gamma}{N_{ph}^2\rho_{e-ph}[\rho_{e-ph}(1-\alpha)(1-f)+f/2]}.
\end{equation}
\vspace{0.2cm}
In the above, we have taken a uniform limit for the energy partition: $p(z) = 1$. 

Here, we have defined $\langle e_e \rangle \equiv \langle E_e \rangle /N_e \equiv $
mean electron energy; $\langle e_{ph} \rangle \equiv \langle E_{ph} \rangle /N_{ph} 
\equiv $ mean phonon energy $(= k_B T_B)$; and $\rho_{e-ph}=N_e/N_{ph}\equiv$ electron-to-
phonon number ratio. It is to be noted that in the limit $\alpha = 1$ ({\it i.e.,} for elastic 
electron-electron collisions as is usually expected for an electronic system unlike the case of the granular gas), and $g_{ex}=0$ ({\it i.e.,} no photo-excitation), we recover $\langle e_e \rangle=\langle e_{ph}\rangle$,
 {\it i.e.,} the electrons and the phonons are at the same temperature, as is physically expected under equilibrium conditions. In general, however, the mean electron energy in the steady state 
is not the same as the mean phonon energy, and the former depends on the excitation rate (the drive $g_{ex}$).  In the following, we will consider other choices for $p(z)$.
\vspace{0.40cm}

\noindent
{\bf Case 1 (\it Extreme Partition Limit)}: $p(z) = \frac{1}{2}(\delta(z) + \delta(z-1))$ 

\noindent 
Analytic form for $\tilde{n}_e(s)$ in the steady state can be obtained for the particular
case of elastic scattering, ($\alpha = 1$), and for the partitioning function 
$p(z)=\frac{1}{2}\delta(z)+\frac{1}{2}\delta(z-1)$.  It can be easily verified that for this insertion of $p(z)$, the Eqs. (5) and (6) remain unchanged. So, with this choice, Eq. (3) becomes 
\begin{eqnarray} 
&&\tilde{n}_{e}(s)[(1-f)N_e+fN_{ph}]=(1-f)\int_0^1(\frac{1}{2}\delta(z)+\frac{1}{2}\delta(z-1
)) dz\tilde{n}_{e}^2(zs)+f\int_0^1 (\frac{1}{2}\delta(z) \nonumber \\
&&+\frac{1}{2}\delta(z-1))dz\tilde{n}_{e}(zs)\tilde{n}_{ph}(zs)
+g_{ex}[e^{-sE_{ex}}-e^{-sE_d}]/\Gamma.
\end{eqnarray}
This can be readily solved to give,
\begin{equation}
\tilde{n}_{e}(s)=\frac{2[(1-f)N_e+fN_{ph}]-f\tilde{n}_{ph}(s)}{2(1-f)} \pm \frac{\sqrt{
[f\tilde{n}_{ph}(s)-2[(1-f)N_e+fN_{ph}]]^2 - 4(1-f)C}}{2(1-f)}.
\end{equation}
with 
\[C = (1-f)N_e^2+fN_eN_{ph} +2g_{ex}(e^{-s E_{ex}}-e^{-s E_d})/\Gamma.\]
We have to choose the $-ve$ sign in the above solution so as to satisfy 
$\tilde{n}_{e}(0)=N_e,\,\,\tilde{n}_{ph}(0)=N_{ph}$.
Here, we have assumed that the bath phonons obey the equilibrium Boltzmann distribution 
(${\it i.e.,}\,\, \tilde{n}_{ph}(s)=N_{ph}/(1+\langle E_{ph}\rangle s))$. It is to be noted here that $\tilde{n}_{e}(s)$ is not zero in the limit $s\rightarrow \infty$. This is due to the fact 
that there is a delta-function peak of strength 
\begin{equation}
\eta = N_{ph}\sqrt{\rho_{e-ph} + \frac{f}{1-f}}\left[\sqrt{\rho_{e-ph} + \frac{f}{1-f}} -
\sqrt{\frac{f}{1-f}} \right]
\end{equation}
at energy $E = 0$. The strength of the delta function reduces to $N_{ph}\frac{1}
{1+\sqrt{f}}$ when there are equal number of electrons and phonons in the sample. It is to be noted that we have here three macroscopic unknown quantities, $N_e, <E_e>$, and the strength $\eta$ of the delta-peak at $E = 0$, while there are only two equations namely, Eq. (6) and Eq. (9). It is, therefore, necessary to solve Eq.(8) for the full distribution ${\tilde{n}_e}(s)$ in order to determine these macroscopic quantities. We have, however, a simpler case  when the depletion by recombination is at the bottom of the band, {\it i.e.,} $E_d$ = 0. In this case, the last term in the rate equation (1) becomes $-g_d \delta(E) n_e(E)$.  Thus, for a delta-function peak in $n_e(E)$ of strength $\eta$ at $E = 0$, this term becomes $-\eta g_d\delta^2(E)$. Recalling that $\delta^2(E) = \xi\delta(E)/2\pi$ with $\xi$ an infinitely large conjugate interval, we can redefine a renormalized depletion rate constant $g_d\xi/2\pi = G_d$, giving the depletion term as $- \eta G_d \delta(E)$. This enables us to determine the condensate strength $\eta$ directly in terms of the photo-excitation rate $g_{ex}$.
\vspace{0.40cm}

\noindent
{\bf Case II  (Equipartition Limit)}: $p(z) = \delta(z-\frac{1}{2})$

\noindent
Finally, we solve Eq.(3) for the steady-state condition for elastic collisions ($\alpha = 1$) with equipartition of energy $(z=1/2)$ between the colliding particles. In this case also, it can be easily verified that with this insertion of $z=1/2$, the Eqs.(5) and (6) remain unchanged.
 For this special case, Eq.(3) reduces to 
\begin{equation}
\tilde{n}_{e}(s)[(1-f)N_e+fN_{ph}]=(1-f)\tilde{n}_{e}^2(s/2)+f\tilde{n}_{e}(s/2)
\tilde{n}_{ph}(s/2)+\frac{g_{ex}}{\Gamma}(e^{-sE_{ex}}-e^{-sE_d}).
\end{equation}
Now, in the limit of small $s$, the $\tilde{n}_{e}(s)$ can be written as,
\begin{eqnarray}
&&\tilde{n}_{e}(s)=\int_0^\infty e^{-sE}n_{e}(E)dE \simeq \int_0^\infty (1-(s/2)E...)
e^{-(s/2)E}n_{e}(E)dE \nonumber \\
&&\simeq \tilde{n}_{e}(s/2)-(s/2)\langle E_{e} \rangle-(s/2)^2\langle E_{e}^2\rangle.
\end{eqnarray}
Using Eqs. (10) and (11), we obtain the electron distribution function
in the $s$-domain as,
\begin{equation}
\tilde{n}_{e}(s/2)\simeq \frac{(1-f)N_e +fN_{ph}-f\tilde{n}_{ph}(s/2)}{2(1-f)}+
\frac{\sqrt{[(1-f)N_e +fN_{ph}-f\tilde{n}_{ph}(s/2)]^2-4(1-f)C_1}}{2(1-f)}
\end{equation}
with
\[C_1 = s\langle E_{e} \rangle [(1-f)N_e+fN_{ph}]/2+g_{ex}(e^{-sE_{ex}}-e^{-sE_d})/\Gamma\].

To invert the above expression for the electron-number density distribution from the $s$-domain to the $E$-domain, we have carried out the small-$s$ analysis. Equation (12) can then be written as
\begin{equation}
\tilde{n}_{e}(s/2)\simeq \frac{2N_e}{\langle E_{ph} \rangle} + \left[1- \frac{2}{\langle 
E_{ph} \rangle}\left(\frac{1}{s+2/\langle E_{ph} \rangle}\right) \right] \left[\frac{[(1-f)N_e +fN_{ph}]
\langle E_{ph} \rangle N_e -2k_1}{(1-f)\langle E_{ph} \rangle N_e}\right],
\end{equation}
with
\[k_1=\langle E_{e} \rangle [(1-f)N_e+fN_{ph}]/2+g_{ex}(E_{ex}-E_d)/\Gamma.\]
We finally obtain
\begin{eqnarray}
&&n_{e}(E) \simeq \left[N_e + \frac{f}{1-f}N_{ph}-\frac{\langle E_{e} \rangle}{\langle E_{ph} \rangle}
\left(1+\frac{f}{(1-f)\rho_{e-ph}}\right)+\frac{2g_{ex}(E_{ex}-E_d)}{(1-f)\Gamma \langle E_{ph}
 \rangle N_e}\right] \nonumber \\
&&\times\left\{\delta(E) -\frac{4}{\langle E_{ph} \rangle}e^{-4\frac{E}{\langle E_{ph} 
\rangle}}\right\}+\frac{4N_e}{\langle E_{ph} \rangle}e^{-4\frac{E}{\langle E_{ph} \rangle}}.
\end{eqnarray}  
Clearly, it has two parts $--$ the Boltzmann distribution at temperature $T$,  and a delta-function condensate at energy $E = 0$.
\vspace{0.40cm}

\noindent
{\bf DISCUSSION}\\
We have treated here the problem of energy-distribution of photo excited electrons in a semiconducting sample as a generalization of the model for dissipative granular gas driven far from equilibrium.  An interesting feature of the above non-equilibrium distribution of the cw photo-excited electrons is the delta-function peak (classical condensate) appearing at the bottom of the conduction band. It should be possible to probe this steady-state feature through a pulsed pump-probe experiment combined with the cw excitation. Of course, for the full electron-distribution function, the Laplace transformed $n_e(s)$ has to be inverted numerically. It is to be noted that we have assumed the sample to be photo-excited homogeneously ({\it i.e.,} only spectral but no spatial diffusion). This is clearly relevant to nanometric-scale samples. As for the dissipative feature ($\alpha < 1$) of the binary electron-electron collisions, it can possibly derive from the dissipative polarization of the dielectric medium that mediates the electron-electron Coulombic interaction [6].

\end{document}